# On the Non Specific Nature of Classical Turbulence Statistics

Trinh, Khanh Tuoc

Institute of Food Nutrition and Human Health

Massey University, New Zealand

*K.T.Trinh@massey.ac.nz*

## Abstract

The classical statistics of turbulence are shown to be not specific to turbulence and can be derived from a solution for recurring unsteady state viscous flow. Care must be exercised in using them to make deductions about turbulence structures and mechanisms. The conditionally averaged statistics, particularly involving the velocities of the ejections in the burst phase, are more distinctive of turbulence.

## 1      Introduction

Turbulence is a complex time dependent three-dimensional motion widely believed to be governed by equations[1] established independently by Navier and Stokes more than 150 years ago

$$\frac{\partial}{\partial t}(\rho u_i) = -\frac{\partial}{\partial x_i}(\rho u_i u_j) - \frac{\partial}{\partial x_i}p - \frac{\partial}{\partial x_i}\tau_{ij} + \rho g_i \qquad (1)$$

and the equation of continuity

$$\frac{\partial}{\partial x_i}(\rho u_i) = 0 \qquad (2)$$

The omnipresence of turbulence in many areas of interest such as aerodynamics, meteorology and process engineering, to name only a few, has nonetheless led to a voluminous literature.

---

[1] The suffices i and j in this paper refer to standard vector notation.

Most of the interest in turbulence modelling from a practical engineering view point was originally based on the time averaged parameters of the steady state flow field. Reynolds (1895) has proposed that the instantaneous velocity $u_i$ at any point may be decomposed into a long-time average value $U_i$ and a fluctuating term $U'_i$.

$$u_i = U_i + U'_i \tag{3}$$

with

$$U_i = \lim_{t \to \infty} \int_0^t u_i \, dt \tag{3.1}$$

$$\int_0^\infty U'_i \, dt = 0 \tag{3.2}$$

For simplicity, we will consider the case when
    1. The pressure gradient and the body forces can be neglected
    2. The fluid is incompressible ($\rho$ is constant).

Substituting equation (3) into (1) and taking account of the continuity equation (2) gives:

$$U_i \frac{\partial U_j}{\partial x_j} = \nu \frac{\partial^2 U_i}{\partial x_i} - \frac{\partial \overline{U'_i U'_j}}{\partial x_i} \tag{4}$$

These are the famous Reynolds equations (Schlichting, (1960), p. 529) also called Reynolds-Averaged-Navier-Stokes equations RANS (Gatski and Rumsey, 2002, Hanjalić and Jakirlić, 2002). The long-time-averaged products $\overline{U'_i U'_j}$ arise from the non-linearity of the Navier-Stokes equations. They have the dimensions of stress and are known as the Reynolds stresses. They are absent in steady laminar flow and form the distinguishing features of turbulence.

The NS equations and the equation of continuity form a closed set that can be solved in principle, even though no general solution has been obtained in the last 160 years because of the great difficulties arising from the non linear terms. When Reynolds averaged the NS equations a degree of freedom is lost and there is no longer sufficient information to solve this new set of equations. This is the famous closure problem in turbulence. It is solved currently by formulating the Reynolds stresses with empirical or semi theoretical models.

Modelling the Reynolds stresses allows us to predict the friction drag and flow patterns but do not give fundamental insight into the structure of turbulence. Taylor (1935) introduced the statistical theory of turbulence to analyse the instantaneous fluctuations of the measured parameters, usually the local velocity and pressure. A second newer approach is the actual observations and numerical simulations of coherent fluid structures embedded in the flow field. These studies have resulted in a voluminous literature that is well summarised in many textbooks on turbulence both old and new e.g. (Hinze, 1959, Lesieur, 2008, McComb, 1991, Schlichting, 1960, Tsinober, 2001).

This paper discusses the prediction and significance of the statistics used in the classical literature to characterise turbulence.

## 2      Theory

In 1967, Kline et al. (1967) reported their now classic hydrogen bubble visualisation of events near the wall and ushered in a new area of turbulence research based on the so-called coherent structures. Despite the prevalence of viscous diffusion of momentum close to the wall, the flow was not laminar in the steady-state sense envisaged by Prandtl (1935). Instead the region near the wall was the most active in the entire flow field. In plan view, Kline et al. observed a typical pattern of alternate low– and high-speed streaks. The low-speed streaks tended to lift, oscillate and eventually eject away from the wall in a violent burst. In side view, they recorded periodic inrushes of fast fluid from the outer region towards the wall. This fluid was then deflected into a vortical sweep along the wall. The low-speed streaks appeared to be made up of fluid underneath the travelling vortex. The bursts can be compared to jets of fluids that penetrate into the main flow, and get slowly deflected until they become eventually aligned with the direction of the main flow. This situation is found in studies of a vortex moving above a wall e.g. (Walker, 1978, Peridier et al., 1991, Smith et al., 1991, Suponitsky et al.,

2005) and results in the growth of the fluctuations and eventually eruption of the low-speed fluid beneath the vortex. The inrush-sweep-burst cycle is now regarded as central to the production of turbulence near a wall.

In another publication (Trinh, 2009a) it has been shown that the wall layer process can be better analysed by decomposition the local instantaneous velocity into four components rather than two as in Reynolds original work. The sweep phase can be modelled as a Kolmogorov flow (Obukhov, 1983) a simple two dimensional sinusoidal flow, or better still analysed with techniques borrowed from laminar oscillating flow (Trinh, 2009b, Trinh, 1992).

The instantaneous velocity in the sweep phase with timescale $t_v$ may be decomposed as:

$$u_i = \tilde{u}_i + u'_i \tag{5}$$

where $\tilde{u}_i$ is the smoothed phase velocity and $u'_i$ fast fluctuations of period $t_f$.

Comparing equations (3) and (5) shows that

$$U_i = \frac{1}{t_v} \int_0^{t_v} \tilde{u}_i \, dt \tag{6}$$

and

$$U'_i = \tilde{U}'_i + u'_i \tag{7}$$

where

$$\tilde{U}'_i = \tilde{u}_i - U_i \tag{8}$$

then

$$u_i = U_i + \tilde{U}'_i + u'_i \tag{9}$$

The traditional approach to analyse unsteady oscillating flows is by a method of successive approximations (Schlichting, 1979, Tetlionis, 1981). The dimensionless parameter defining these successive approximations is

$$\varepsilon = \frac{U_e}{L\omega} \tag{10}$$

where $U_e$ is the local mainstream velocity and L is a characteristic dimension of the body. The smoothed velocity $\tilde{u}_i$ is given by the solution of order $\varepsilon^0$ which applies when $\varepsilon \ll 1$. The governing equation (Einstein and Li, 1956, Hanratty, 1956, Meek and Baer, 1970, Trinh, 2009b) is a subset of the NS equations

$$\frac{\partial \tilde{u}}{\partial t} = \nu \frac{\partial^2 \tilde{u}}{\partial y^2} \tag{11}$$

where $\tilde{u}$ refers here to the smoothed velocity $\tilde{u}_i$ in the $x$ direction. It does not require that there are no velocity fluctuations, only that they are small enough for their effect on the smoothed phase velocity $\tilde{u}$ to be negligible. Stokes (1851) has solved this equation for the conditions:

| | | | |
|---|---|---|---|
| IC | $t = 0$ | all y | $\tilde{u} = U_\nu$ |
| BC1 | $t > 0$ | $y = 0$ | $u = 0$ |
| BC2 | $t > 0$ | $y = \infty$ | $\tilde{u} = U_\nu$ |

where $U_\nu$ is the approach velocity for this sub-boundary layer. The velocity at any time t after the start of a period is given by:

$$\frac{\tilde{u}}{U_\nu} = \text{erf}(\eta_s) \tag{12}$$

where $\eta_s = \frac{y}{\sqrt{4\nu t}}$

The average wall-shear stress is

$$\tau_w = \frac{\mu U_\nu}{t_\nu} \int_0^{t_\nu} \left(\frac{\partial \tilde{u}}{\partial y}\right)_{y=0} dt = \frac{\mu U_\nu}{t_\nu \sqrt{\pi}} \int_0^{t_\nu} \frac{1}{\sqrt{\nu t}} dt \tag{13}$$

Equation (16) may be rearranged as

$$t_\nu^+ = \frac{2}{\sqrt{\pi}} U_\nu^+ \tag{14}$$

The time-averaged velocity profile near the wall may be obtained by rearranging equation (12) as

$$\frac{U^+}{U_\nu^+} = \int \text{erf}\left(\frac{y^+}{4U_\nu^+ \sqrt{t/t_\nu}}\right) d\left(\frac{t}{t_\nu}\right) \tag{15}$$

Equation (15) applies up to the edge of the wall layer where $u/U_\nu = 0.99$, which corresponds to $y = \delta_\nu$ and $\eta_s = 1.87$. Substituting these values into equation (12) gives

$$\delta_\nu^+ = 4.16 U_\nu^+ \tag{16}$$

Back-substitution of equation (16) into (14) gives

$$\delta_v^+ = 3.78 t_v^+ \qquad (17)$$

where the velocity, period and normal distance have been normalised with the wall parameters $v$ the kinematic viscosity and $u_* = \sqrt{\tau_w/\rho}$ the friction velocity, $\tau_w$ the time averaged wall shear stress and $\rho$ the density.

As the vortex moves along the wall, the magnitude of the velocity fluctuations grows ($\varepsilon$ increases) according to well-known analyses of stability of laminar flows e.g. (Dryden, 1934, Dryden, 1936, Schiller, 1922, Schlichting, 1932, Schlichting, 1933, Schlichting, 1935, Schubauer and Skramstad, 1943, Tollmien, 1929). We then switch to a second approximation of order $\varepsilon$. We may average the Navier-Stokes equations over the period $t_f$ of the fast fluctuations. Bird, Stewart and Lightfoot (1960, p. 158) give the results as

$$\frac{\partial(\rho \tilde{u}_i)}{\partial t} = -\frac{\partial p}{\partial x_i} + \mu \frac{\partial^2 \tilde{u}_i}{\partial x_j^2} - \frac{\partial \tilde{u}_i \tilde{u}_j}{\partial x_j} - \frac{\partial \overline{u'_i u'_j}}{\partial x_j} \qquad (18)$$

Equation (18) defines a second set of Reynolds stresses $\overline{u'_i u'_j}$ which we will call "fast" Reynolds stresses to differentiate them from the standard Reynolds stresses $\overline{U'_i U'_j}$. Within a period of the wall layer process $t_v$, the smoothed velocity $\tilde{u}_i$ varies slowly with time but the fluctuations $u'_i$ may be assumed to be periodic with a timescale $t_f$. In the particular case of steady laminar flow, $\tilde{u}_i = U_i$ and $\tilde{U}'_i = 0$. We may write the fast fluctuations in the form

$$u'_i = u_{0,i} \left( e^{i\omega t} + e^{-i\omega t} \right) \qquad (19)$$

The fast Reynolds stresses $u'_i u'_j$ become

$$u'_i u'_j = u_{0,i} u_{0,j} (e^{2i\omega t} + e^{-2i\omega t}) + 2 u_{0,i} u_{0,j} \qquad (20)$$

Equation (20) shows that the fluctuating periodic motion $u'_i$ generates two components of the "fast" Reynolds stresses: one is oscillating and cancels out upon long-time-averaging, the other, $u_{0,i} u_{0,j}$ is persistent in the sense that it does not depend on the period $t_f$. The term $u_{0,i} u_{0,j}$ indicates the startling possibility that a purely oscillating motion can generate a steady motion which is not aligned in the direction of the oscillations. The qualification steady must be understood as independent of the frequency $\omega$ of the fast fluctuations. If the flow is averaged over a longer time than the period $t_v$ of the bursting process, the term $u_{0,i}u_{0,j}$ must be understood as transient but non-oscillating. This term indicates the presence of transient shear layers embedded in turbulent flow fields and not aligned in the stream wise direction similar to those associated with the streaming flow in oscillating laminar boundary layers. Thus the instantaneous velocity in terms of at least 4 components:

$$u_i = U_i + \tilde{U}'_i + u'_{i_i}(\omega t) + u_{i,st} \tag{21}$$

where $u_{i,st}$ is the velocity of the streaming flow or ejection of wall fluid.

It will be noted from this analysis made that the solution of order $\varepsilon^0$ is independent of the solution of order $\varepsilon$ because it applies to the sweep phase before the streaming flow has grown significantly.. The sweep phase lasts much longer than the bursting phase e.g. (Walker et al., 1989) and dominates the average velocity distribution in the wall layer.

## 3    The time-averaged velocity profile of the wall layer

Einstein and Li (op.cit.) and Hanratty (op.cit.) have applied equation (15) between the wall and the edge of the buffer layer (Karman, 1934), where $U_b^+ = 13.5$ and $\delta_b^+ = 30$ and obtained good agreement with measured velocity profiles. Meek and Baer (op.cit.) matched equation (12) with Prandtl's law of the wall (Prandtl, 1935)

$$U^+ = 2.5\ln y^+ + 5.5 \tag{22}$$

and obtained $U_v^+ = 14.9$, $\delta_v^+ = 64$. Substituting this new criterion into equation (15) gave good predictions for the time-averaged velocity over the whole wall layer, $0 < y^+ < 64$. Meek and Baer also showed that equation (14) gave a very good prediction of the time scale of the wall layer process as shown in Figure 1.

## 4    Statistics of the solution of order $\varepsilon^0$

We now calculate the statistics of the sweep phase by using the solution of order $\varepsilon^0$ and compare them with the classical statistics obtained from the measurements of the instantaneous velocity in turbulent flows.

## 4.1 Probability density distribution of the streamwise velocity

Eckelmann (1974) made measurements of the probability density distribution (pdf) of the instantaneous velocity in an oil channel with an average velocity of 22.5 cm/s by positioning an array of probes at various distances $y^+$ from the wall and sampling up to $8.10^5$ measurements for the number of occurrences in a given velocity interval.

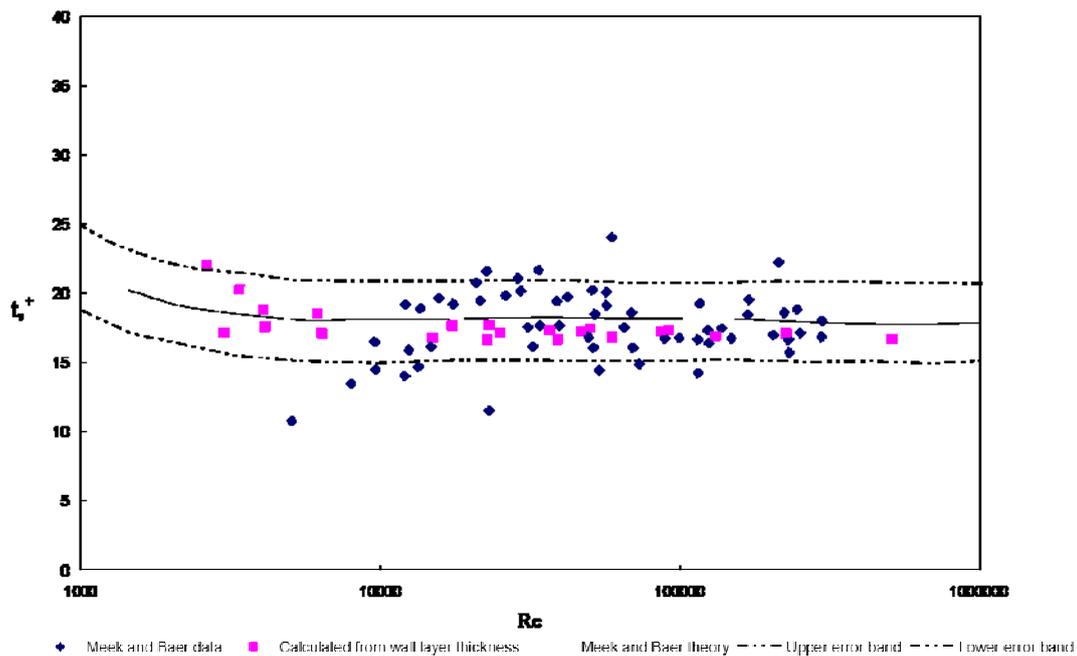

Figure 1. The time scale of the wall layer process according to Meek and Baer (1970)

The number of observations which occur in each of the velocity intervals divided by the total number of measurements and by the interval width (0.28 cm/s) gives an approximate measure of the probability density P(u) of the streamwise velocity (Figure 2a).

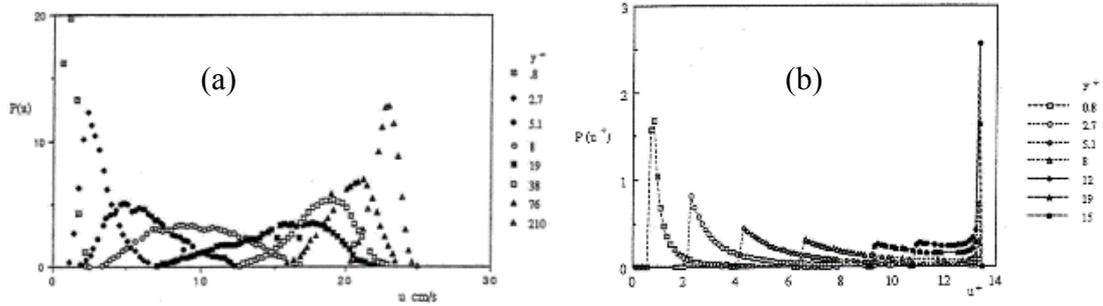

Figure 2. (a) Probability density function educed from the Stokes solution1, (b) Data of Eckelman (1974). Reproduced from Trinh (1992, 2005)

Eckelmann's experiment was simulated on a personal computer by generating 2000 point values of the smoothed velocity using equation (12) for each position $y^+$ measured by Eckelmann over regular intervals of $t/t_v$ spanning from 0 to 1. For each value of $y^+$, the generated population was sampled for rate of occurrence in a given interval of normalised streamwise velocity. The probability density was obtained by dividing the number of occurrences by 2000 and by the width of the interval. Numerical comparison with the data of Eckelmann can be made by multiplying the peak value of the normalised velocities with the friction velocity of 1.15 cm/s prevalent in his experiment.

Two sets of scaling parameters, $U_v^+$ and $\delta_v^+$ were tested. These were

1) The scales at the edge of the wall layer ($U_v^+ = 17$, $\delta_v^+ = 70$) determined from the intersection of equation (16) with the experimental velocity profile of Eckelmann. This technique for determining the edge of the Stokes/wall layer is described in more detail elsewhere (1992, Trinh, 2009b).
2) The scales at the edge of the buffer layer ($U_v^+ = 14$, $\delta_v^+ = 30$) defined by von Karman (1934)..

The pdf obtained from both these sets of scaling parameters had similar shape but the second set, shown in Figure 2b, gave more accurate estimates of the position of the peaks in the pdf. This is because the ejections, which are not accounted for by equation (12), also contribute to the experimental velocity distribution in the region $30 < y^+ < 70$.

Two basic differences between the simulation and the experiment of Eckelmann must be stressed:

1.  Eckelmann's velocity population spanned the whole range of possible values across the channel, *i.e.* from 0 to $U_m$ at the axis. In this simulation, velocities were limited to $U_v^+$ ($\approx$ 2/3 $U_m^+$), that at the edge of the buffer layer. Thus the generated population is an underestimate of the real population over which the instantaneous velocity was sampled. This condition makes the estimated probability density distribution peak higher than the measured value.
2.  The Stokes solution assumes a uniform bonding approach velocity U. The simulation therefore forces the velocity at the edge of the wall layer to a probability of 1. In reality the velocity fluctuations persist right to the channel axis.

Despite these restrictions, the generated probability distribution shows remarkable similarity to that measured. For each $y^+$, the peak on the two figures 2a and 2b occurs at the same values of $U^+$. This indicates that the Stokes solution correctly estimates the value of the dominant wavelengths. As the distance from the wall $y^+$ increases from 0 to 13, the relative size of the peaks diminish in the proportions measured. The probability density distributions (pdf) become more widely spread.

For values of $y^+$ below 12, the maximum in the profile of fluctuating velocity, the curve is skewed to the left. For the remainder it is skewed to the right. The pdf for $y^+ = 13$ has not been measured by Eckelmann but can be shown to be a minimum in this simulation. This trend can be observed on the predicted plot but a definite bias to the left is strongly apparent even though some skewing to the right occurs near the edge of the buffer layer. This is probably because the model assumes that at the end of the period the initial uniform flow profile is restored abruptly. This is not borne out in reality.

The agreement between Figure 2 (a) and (b) is better for values of $y^+ < 12$ than in the remainder of the buffer layer. This suggests that the Stokes solution models best the front portion of the sweeps where the Reynolds stresses are minimal and the disturbing influence of the ejections is least.

**4.2    The dominant velocity**

When one overlays the values of the most probable velocities at various position $y^+$

obtained by Eckelmann with the values predicted in Figure 2 (a), the correspondence is remarkable (Figure 3). The distribution of the most probable velocity also coincided with the profile of time averaged velocity in the wall layer.

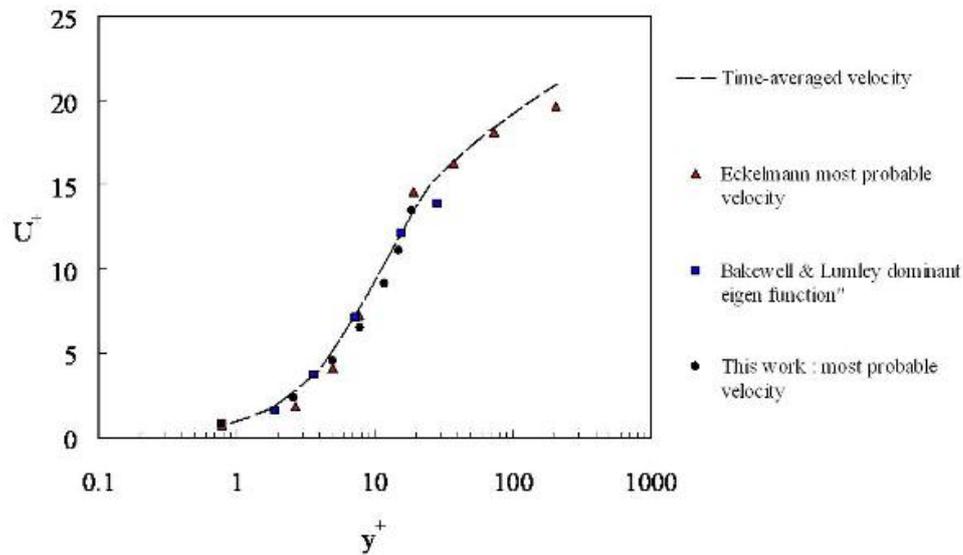

Figure 3 Dominant velocities compared to time averaged velocity profile. Data from (Bakewell & Lumley, 1967; Eckelmann, 1979; Trinh, 1992)

Bakewell & Lumley (1967) made an orthogonal decomposition of the instantaneous velocity traces, a tool that has also been used successfully in other flow geometries, e.g. by Takeda (1999) for the study of turbulent Taylor vortex flow. Such decompositions were supposed to identify the most important velocity fluctuations in turbulent flow fields. The dominant Eigen function in Bakewell and Lumley's orthogonal decomposition also agreed with the profiles of the most probable velocity predicted and the time averaged velocity measured by many authors e.g. Laufer (1954).

Considering the crudity of the model, the agreement is already enlightening. A better model is obtained by noting that the ejections result in a non-uniform velocity outside the wall layer and the approach velocity in the Stokes solution needs to be modified accordingly for a more rigorous analysis.

## 4.3 The moving front of turbulence

The Stokes solution cannot show how the unsteady viscous state sub-boundary layer behaves in the x direction. A picture can be obtained by using a time-space transformation (Trinh and Keey, 1992a, Trinh and Keey, 1992b, Trinh, 2009b). Trinh and Keey showed that the Stokes solution can be transformed exactly into the Blasius solution for a laminar boundary layer on a flat plate (Blasius, 1908) with an extended form of Taylor's hypothesis which yields

$$\delta_v = 4.96 \sqrt{\frac{x_v \nu}{U_v}} \qquad (23)$$

In later papers Trinh (2002, 2009b, 2010) showed that the velocity derivative in equation (12) must be interpreted as a Lagrangian derivative along the path of diffusion of viscous momentum.

Using an array of hot wires and wall shear stress probes, Kreplin & Eckelmann (1979) made correlations of the wall shear and the local instantaneous velocity within the wall layer. The long time correlations exhibited clear peaks.

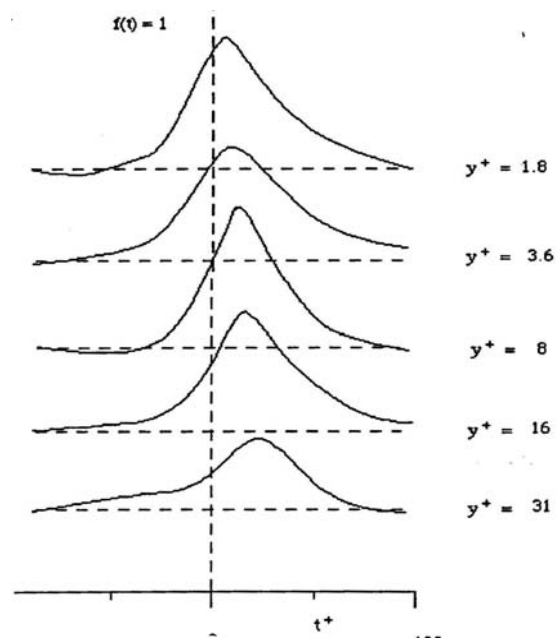

Figure 4. Shift of the autocorrelation peak. From Kreplin & Eckelmann (1979).

They transformed the time shift in the peaks of their correlations into spacing in the x direction by multiplying with a convection velocity measured by wall probes spaced in the x direction. Kreplin and Eckelmann deduced from these measurements a "moving front of turbulence' in the wall layer. The probes used by Kreplin and Eckelmann are fixed thus the Lagrangian solution of equation (12) must be transposed into an Eulerian context for suitable comparison. We assume that slow speed streaks of all ages t (and therefore lengths x) have an equal probability of passing a fixed probe in the flow field. Then a statistical average for the sub-boundary layer thickness is obtained by averaging over all x to obtain a

$$\delta_b = 3.31\sqrt{\frac{x_\nu \nu}{U_\nu}} \tag{24}$$

which may be rearranged as (Trinh 1992)

$$\delta_b^+ = 1.28\sqrt{x_\nu^+} \tag{25}$$

Equation (25) fits the 'moving front of turbulence' of Kreplin and Eckelmann perfectly as shown in Figure 5.

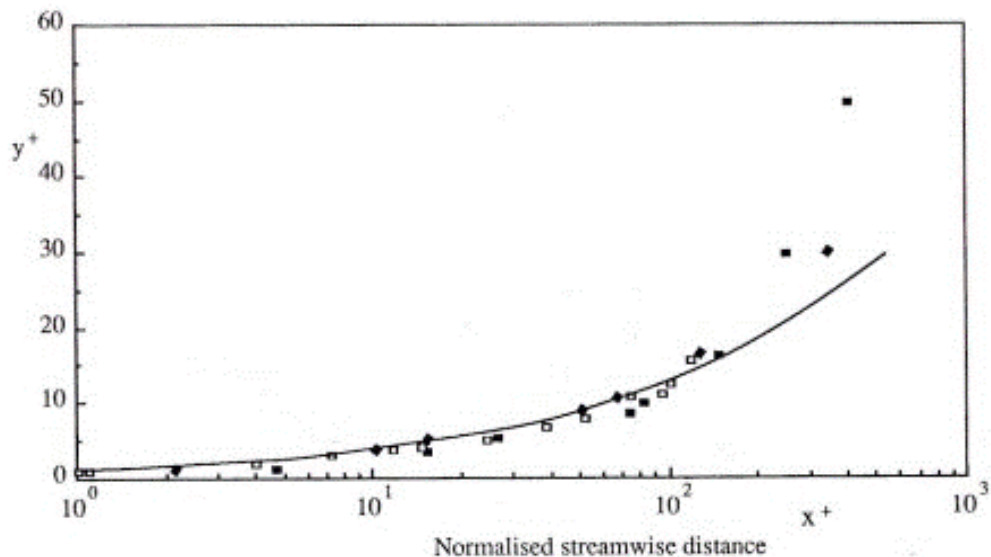

Figure 5 "Moving front of turbulence" and statistically averaged edge of the solution of order $\varepsilon^0$ using equation (25). Data of Kreplin and Eckelmann (1979).

Since the low-speed streak represents fluid just below the moving longitudinal vortex above the wall, Kreplin and Eckelmann's "moving front of turbulence" is better interpreted as the path of the vortex. Thus the shape of the hairpin vortex can be explained by two completely separate events. The legs of the vortex are shaped in the sweep phase because the viscous sub-boundary layer that is induced under the travelling vortex growth in thickness as the diffusion of viscous momentum penetrates into the main flow pushing the vortex further away. Then in the following bursting phase, the streaming ejection, whose path is well captured by the log-law (Trinh, 2009b) p. 20 lifts the head to a much steeper angle.

### 4.4 The fluctuating velocity

The r.m.s. fluctuating velocity at any point $y^+$ is given by sampling over the distribution of low-speed streaks of all ages, using:

$$\frac{\sqrt{\overline{u'^2}}}{u_*} = \frac{\int_0^{t_v^+} \sqrt{(\tilde{u}^+ - U^+)^2}\, dt^+}{t_v^+} \tag{26}$$

where the smoothed phase velocity is calculated from equation (12) and the long time average velocity from equation (15).

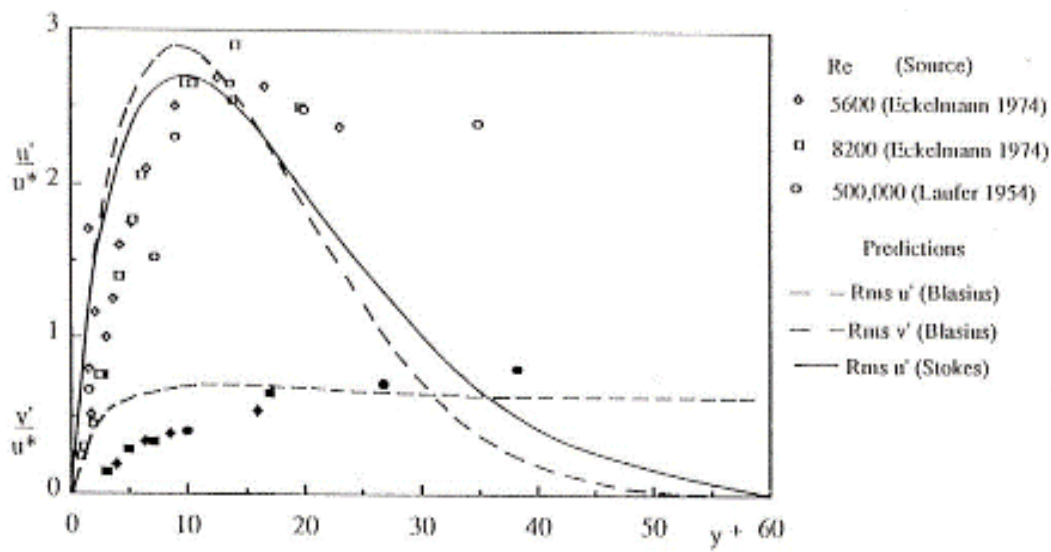

Figure 6. Longitudinal and normal fluctuating velocities near the wall at different Reynolds numbers and predictions from the Stokes solution1 and Blasius solution.

The trends compare well with the measurements of Eckelmann (1974) and Laufer (1954), in Figure 6 but the predicted peak occurs sooner than in the experimental data and the predicted fluctuating velocity in falls off more rapidly. This is because the simple model proposed has assumed a constant approach velocity $U_v^+$ at the edge of the wall layer whereas it is in reality fluctuating. A better model would use an approach velocity consisting of a time averaged value $U_v^+$ and a fluctuating component $u_v'^+$. A similar prediction can be made by averaging the velocity fluctuations from the Blasius solution over all possible lengths $x$ as shown in Figure 6. The advantage of the Blasius solution is to give an estimate of the fluctuating velocity $v'/u_*$ shown in Figure 6.

## 4.5 Correlation function of the wall shear stress

The wall shear stress can be calculated according to the Blasius equation and the correlation coefficient obtained according to the definition

$$f(x) = \frac{\overline{(du/dy)_0 (du/dy)_x}}{\overline{(du/dy)_0^2}} \qquad (27)$$

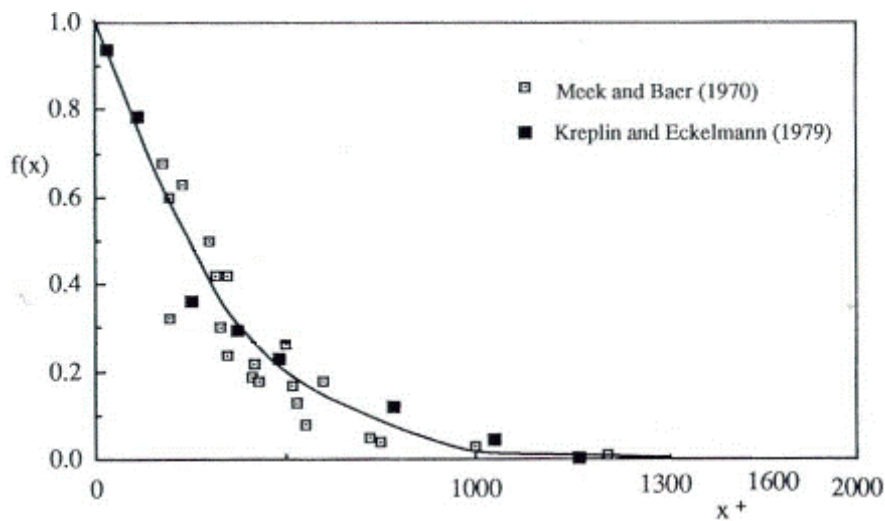

Figure 7 Correlation function for the wall shear stress. Line represents equation (27). Data of Meek and Baer (1970) and Kreplin and Eckelmann (1979).

It is compared with the measurements of Kreplin and Eckelmann and Meek and Baer

(Figure 7), Again agreement is good for the first part of the curve but the correlation function (crf) is overestimated at the tail end of the curve, presumably for the same reason as the fluctuating velocity was underestimated.

### 4.6 The production of turbulence

The production of turbulent energy is defined as

$$P = \overline{U'V'}\frac{dU}{dy} \tag{28}$$

In pipe flow, the local shear stress is (Bird, Stewart and Lightfoot 1960, p. 162)

$$\tau = \tau_v + \overline{U'V'} = \tau_w\left(1 - \frac{y}{R}\right) \tag{29}$$

The viscous shear stress is defined as

$$\tau_v = \mu\frac{dU}{dy} \tag{30}$$

Therefore we may write

$$\frac{\overline{U'V'}}{\tau} = \frac{\tau_t}{\tau} = 1 - \frac{dU^+/dy^+}{1 - y^+/R^+} \tag{31}$$

where $U^+ = U/u_*$. Combining equations (31) and (28) and rearranging gives

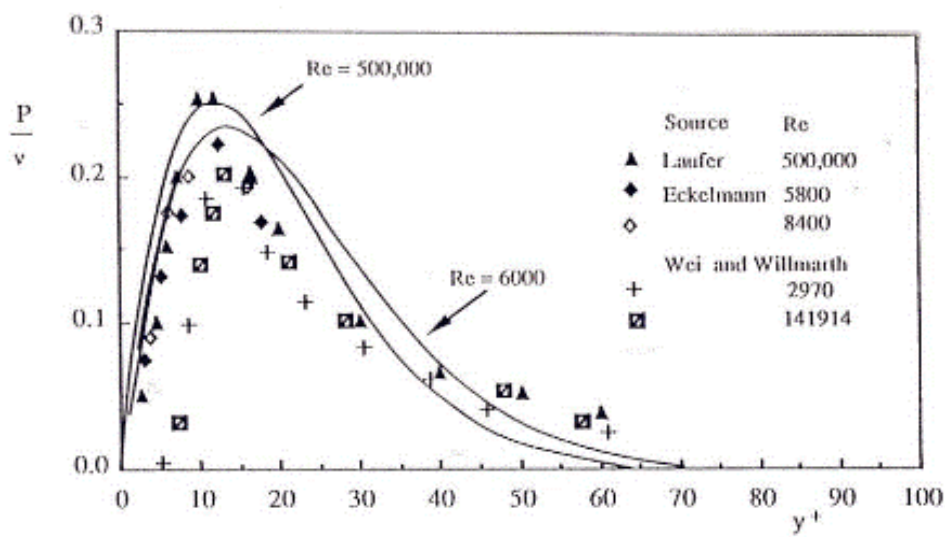

Figure 8 Production of turbulence predicted from the Stokes solution1.

$$\frac{P}{\nu} = \left(1 - \frac{y^+}{R^+}\right)\frac{dU^+}{dy^+} - \left(\frac{dU^+}{dy^+}\right)^2 \tag{32}$$

by using values of the time averaged velocity profile obtained from equation (15), we can estimate the profile of turbulence production as shown in Figure 8 against the data of Laufer (op. cit.) and Eckelmann (op. cit.). Note that normalising the production term P with $U_\nu^+$ and $\delta_\nu^+$ does not result in a single curve because of the presence of the factor $(1-y^+/R^+)$.

**4.7    The energy spectrum and the Kolmogorov scale**

The most powerful data used in turbulence studies is probably the energy spectrum derived from the correlation function, crf. The correlation for the velocity can be obtained from measurements of the instantaneous velocity by, for example, hot anemometers. The auto correlation is given by:

$$f(\tau) = \frac{\overline{u(t)u(t+\tau)}}{\overline{u(t)^2}} \tag{33}$$

where $\tau$ is here a time delay. The two point spatial correlation is

$$f(r) = \frac{\overline{u(x,t)(u(x+r,t)}}{\overline{u(x,t)^2}} \tag{34}$$

It is used to estimate a macroscale $L_x$ (Bradshaw, 1971) which is traditionally interpreted as the size of the eddy passing the probe.

$$L_x = \int f(r)dr \tag{35}$$

and a (Taylor) microscale (Bradshaw, 1971, Lesieur, 2008), an estimate of the size of an eddy where viscous dissipation occurs

$$l_t = \frac{15\overline{u'^2}\nu}{\zeta} \tag{36}$$

where $\zeta$ is the energy per unit volume[2]. This scale can be determined by fitting a vertex parabola to the correlation function. It is traditionally argued that large eddies contain mainly kinetic energy, are unstable and breakdown to smaller and smaller eddies. If the difference between the large and small eddies is large (e.g. at high Reynolds numbers) a wide spectrum of intermediate eddies exist which contain kinetic energy and dissipate little but during the process of degeneration the anisotropic characteristic of the large eddies is lost. Kolmogorov (1941a, 1941b) concluded that the properties of the smallest eddies are statistically independent of the primary eddies and are determined only by the rate of dissipation per unit mass. Thus in a small volume the components of the fluctuating velocity are equal. This situation is called local isotropy and does not require that the bulk stream itself be isotropic.

---

[2] The symbol $\zeta$ is used for the energy per unit volume rather than $\varepsilon$ used in conventional texts on turbulence between the symbol $\varepsilon$ has already been used in with respect to oscillating flow.

Bradshaw (1971) illustrated how vortex stretching at different scales leads to local isotropy with a "family tree" shown in Figure 10. Frisch (Frisch, 1995, Frisch et al., 1978) has illustrated the concept of an energy cascade first postulated by Richardson (1922) by the breakdown of large unstable eddies with no loss of energy until the smallest eddies are reached (Figure 9).

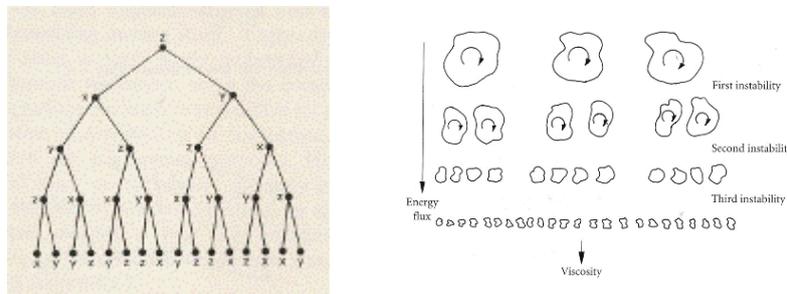

Figure 9. Left: Bradshaw (1971) representation of "family tree for local isotropy. Right: Frisch (1978, 1995) representation of Richardson's (1922) energy cascade.

The energy spectrum describes the flow of energy between different scales and can be obtained from a Fourier transform of the crf

$$\frac{E(k)}{u'^2} = \frac{2}{\pi} \int_{-\infty}^{\infty} f(r) \exp(-ikr) dr \qquad (37)$$

Kolmogorov (op.cit.) argued that the smallest eddies where all the remaining turbulent energy is dissipated must scale with the (kinetic) energy per unit volume $\zeta$ and the viscosity and obtained by dimensional analysis

$$l_k = \frac{\mu^{3/4}}{\rho^{1/2}} \varsigma^{-1/4} \tag{38}$$

Kolmogorov also argued that in the range of unstable eddies, called the inertial sub-range, the only relevant variable for the spectrum is the energy per unit volume and showed by dimensional analysis that

$$E(k) = C(K) k^{-5/3} \varsigma^{2/3} \tag{39}$$

The slope of -5/3 has been widely observed e.g. (Lesieur, 2008).

We can obtain the energy spectrum by taking the Fourier transform of the crf shown in Figure 8 but because the curve predicted by the solution of order $\varepsilon^0$ overestimates the crf at large values of $x^+$ the spectrum obtained will not be accurate for a critical range of wavenumbers of interest. There is another intriguing way of obtaining it from time averaged parameters that allows us to discuss the physical implication of this spectrum.

The pdf predicted in section 4.1 shows that the velocity field passing a probe situated at a distance $y^+$ in composed of a distribution of velocities, each of which must contribute to the total fluid energy at that position. By energy, we mean mainly kinetic energy, which has the dimension of stress. Figure 3 also shows that the contribution at each position $y^+$ reflects the contribution of a dominant Eigen mode. We suspect the reverse to be true: the time-averaged turbulent shear stress at each position $y^+$ may be used as a good estimate of the contribution of the dominant wave length. It is thus edifying to plot out the variation of turbulent stress with distance. In this exercise, we start from the equation

$$\tau = \tau_v + \tau_t \tag{40}$$

The shear stress in a pipe at y is given by

$$\tau = \tau_w (1 - \frac{y}{R}) = \tau_w (1 - \frac{y^+}{R^+}) \tag{41}$$

The laminar contribution can be calculated from Newton's law of viscosity

$$\tau_v = \mu \frac{dU}{dy} \tag{42}$$

Combining equations (40), (41) and (42) gives

$$\frac{\tau_t}{\tau} = 1 - \frac{dU^+/dy^+}{1 - y^+/R^+} \qquad (43)$$

It is useful for comparison purposes to define here a dimensionless wavenumber based on the pipe diameter D and the distance $x^+$ travelled by the dominant Eigen mode and calculated from Taylor's hypothesis and the time scale of the wall layer measured by Meek and Baer (1970)

$$x^+ = U^+ t_v^{+2} \qquad (44)$$

$$k_x D = \frac{2\pi D^+}{x^+} \qquad (45)$$

$U^+$ is the time averaged velocity at position $y^+$, which of course is also equal to the velocity of the dominant Eigen mode. Figure 10 shows a plot of the turbulent shear stress spectrum in a pipe measured by Lawn (1971) at $Re = 9.10^4$. Included are points of $(\tau_t/\tau)/(2\pi)$ against $kR$ calculated with equations (43) and (45) respectively using velocity data of Laufer (op.cit.) and Eckelmann (op.cit.) near the wall. Estimates from the Stokes solution are also included.

The changes in $E(kR)$ and $(\tau_t/\tau)/(2\pi)$ with $kR$ show very similar trends. In particular for wavelengths smaller than $kR \approx 3.5$ both variables tend to level out to a constant value. The drop in of $(\tau_t/\tau)/(2\pi)$ for wavelengths in the range $10 \leq kR \leq 3.5$ is strikingly similar to the drop in $E(kR)$ despite much scatter in the data for $(\tau_t/\tau)/(2\pi)$. This scatter arises from the difficulty of making measurements of velocity very close to the wall. In principle, we can by-pass that difficulty by taking direct measurements of $\overline{u'v'}$ such as those of Eckelmann (1979). These data confirm that equation (43) gives good predictions of $\rho\overline{u'v'}/\tau$ for the range $5 < y^+ < 100$ but since the size of the probe itself is equal to $d_p^+ = 2$, measurements below $y^+ = 5$ show a bias. Calculations from the velocity profile involve numerical differentiation, an inaccurate exercise in itself, but further involve very small differences between two very similar numbers. For example, in order to obtain a value of $E(kR) \approx 10^{-3}$ we must get a slope $dU^+/dy^+$ accurate to four decimal points. Given these difficulties, it is all the more remarkable that the two sets of variables agree so well and that the plot of $(\tau_t/\tau)/(2\pi)$ is able to reproduce so clearly the famous $-5/3$ slope predicted by Kolmogorov.

Yet more physical insight can be obtained by looking at the dissipation scales implied. If we neglect $\tau_t$ in equation (40) we assume that all the energy is dissipative and a rearrangement of equation (42) after integration gives

$$U^+ = y^+ \qquad (46)$$

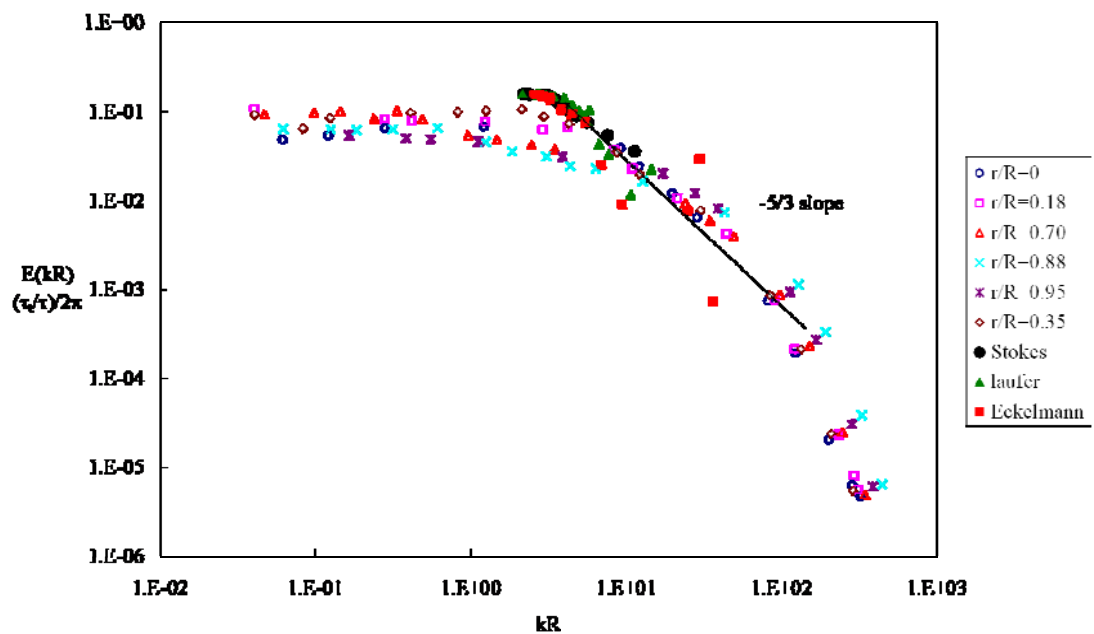

Figure 10. Shear stress and turbulent stress spectra. Spectral data of Lawn (1971), Re=921000, calculated shear stress distributions based on velocity data of Laufer (1954), Eckelmann (1974) and the Stokes solution.

If on the other hand we neglect the term $\tau_v$ in equation (40) and follow Prandtl's derivation we get the log law in equation (22). Thus one might expect that the viscous and kinetic contributions to the local shear stress will be equal at the intersection of equations (22) and (46) which occurs for pipe Reynolds numbers between 6,000 and

$10^6$ at an average value of $y_k^+ = 11.8$ according to experimental data e.g. (Nikuradse, 1932). The reader may verify that indeed at that position $y_k^+ = 11.8$ we obtain the equality $\tau_v = \tau_t$, as shown in (Trinh, 2009b) p.83. That equilibrium between kinetic and viscous energies is characteristic of the equilibrium range of wavenumbers and also represents the defining condition of the Kolmogorov scale equation (38).

It is very interesting to note that the gradient of turbulent stress spectrum in Figure 11 is -5/3 for wavenumbers in the range $4 < kR < 50$ including the point $kR \approx 5$ equivalent to $y_k^+ = 11.8$. We note here that equation (38) implies that the Kolmogorov scale is found at the scale where the turbulent and viscous energies are equal and it is defined by the point where E (or) has dropped to half of its value as low kR which is equal to $kR \approx 5$. The slope of -5/3 applies, in Kolmogorov's argument to the inertial sub-range where the energy spectrum is only dependent on $\zeta$ whereas it is found in the turbulent stress spectrum to straddle the interface of this inertial subrange and the beginning of the dissipation sub range.

## 5    Statistics of the solution of order $\varepsilon$

The generation of most of the classic statistics of turbulence from the solution of order $\varepsilon^0$ which applies equally to steady laminar flow and to the sweep phase of the wall process raises serious questions about using these to differentiate laminar and turbulent flows. While we can agree with Reynolds (1895) that the Reynolds stresses are the distinguishing feature of turbulence, the four component decomposition of the instantaneous velocity in equation (21) highlights the fact that the Reynolds stresses themselves must be decomposed into slow and fast components; the slow components can be obtained from the solution order $\varepsilon^0$; the fast components linked to the velocity of the streaming flow $u_{i,st}$ are the real distinguishing feature of turbulent flow. If the bursting process is suppressed, the flow is simply laminar, even if periodic fluctuations are present.

Johansson, Alfresson, & Kim (1991) analysed the data base provided by the direct numerical simulation DNS of Kim, Moin and Moser (1987) to obtain the

conditionally averaged production of turbulent kinetic energy $\tilde{P}$ which they write as

$$\tilde{P} = \overline{U'V'}\frac{dU}{dy} - \overline{U'V'}\left(\frac{\partial \tilde{U}'}{\partial x} + \frac{\partial \tilde{V}'}{\partial y}\right) - \overline{V'^2}\frac{\partial \tilde{V}'}{\partial y} - \overline{W'^2}\frac{\partial \tilde{W}'}{\partial z} \qquad (47)$$

The first term on the right-hand side of equation (47) is the only one that remains in the long-time averaged sense and is obtained from the solution of order $\varepsilon^0$. It is shown in Figure 11b. The total conditionally averaged production $\tilde{P}$ is substantially higher as seen in Figure 11a. The difference between these two terms is shown in Figure 11c. It points to the existence of an important transient contribution weakly slanted with respect to the wall and which can be attributed to strong gradients in the x- and y- directions of the conditionally averaged streamwise velocity.

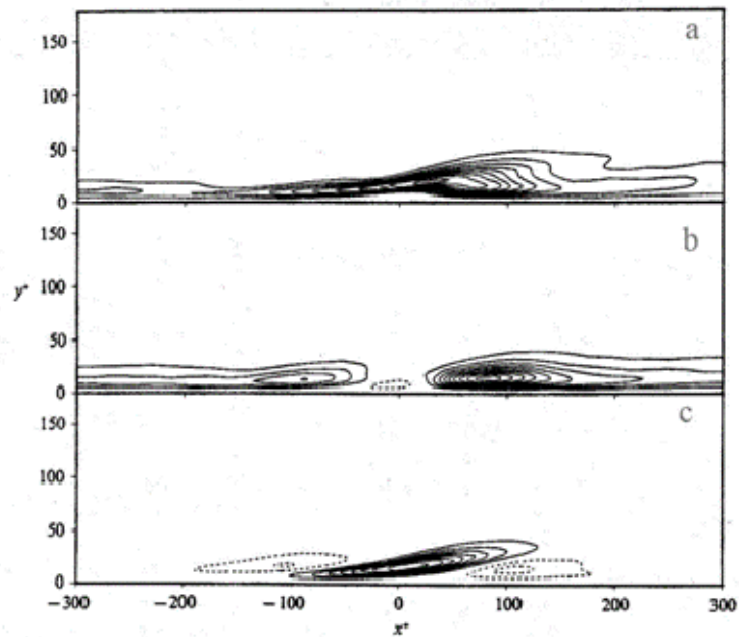

Figure 11 Production of turbulence near the wall. (a) $\tilde{P}$, (b) $\overline{U'V'}(dU/dy)$, (c) $\tilde{P} - \overline{U'V'}(dU/dy)$. After Johansson, Alfresson, & Kim (1991)

Johansson et al. confirm that the Reynolds stresses contribution from the downstream side of the shear layers is spatially spotty but they could follow the associated <U'V'> peaks for distances up to 1000 wall units. Furthermore, they found no signs of oscillatory motions or violent break-up in conjunction with these shear layers which, they believe, indicate a persistent motion of low-speed fluid away from the wall. This is highly suggestive of the ejections widely observed in the wall process. The velocity that Johansson et al. used in equation (47) to calculate the transient contribution to the total conditionally averaged $\widetilde{P}$ are just different symbols for the streaming components of the instantaneous velocity $u_{i,st}$ which can only be obtained from the solutions of order $\varepsilon$ and higher.

## 6   Discussion

It has been known since the ground breaking work of Kline et al. (1967) that much of the turbulent stresses in the wall layer are produced during the bursting phase. The ability of the solution of order $\varepsilon^0$ to predict the long time averaged production of turbulence (section 4.6) simply reflects the fact that the sweep phase lasts much longer than the bursting phase and therefore dominates the statistics of the wall layer (Walker et al., 1989). Nonetheless, it shows that the classical long-time averaged statistics of turbulence are not specific to turbulent flow and only the conditionally averaged statistics that include velocity components from the solution of order $\varepsilon$ and higher are. This puts many of the views based on the classical statistics in a new light.

The Kolmogorov scale has perhaps been the single most useful concept in practical modelling of turbulent flows. It is often described as the size of "smallest, dissipative, turbulent eddies" e.g. Wilson & Thomas (1985) but Kolmogorov himself never argued that the scale bearing his name represented an eddy although he described it as "the scale of the finest pulsations". The mathematical formulation of the Kolmogorov scale in equation (38) is based on his first similarity hypothesis that "the distributions $F_n$ are uniquely determined by the quantities $\nu$ and $\zeta$" and obtained by dimensional analysis. It merely identifies the scale where the (mainly kinetic) energy in the flow is equal to the viscous (dissipation) energy. The point (11.8, 11.8) used by many authors to predict transport rates e.g. (Levich, 1962, Wilson and Thomas, 1985, Metzner and Friend, 1958) is based on the intersection of equations (22) and (48). It is a fictitious point and does not coincide on any real point on any measured velocity profile of turbulent flow. As far as I know there is no published evidence of a "Kolmogorov eddy" either in visual experiments or DNS data bases.

It is well known that the section of the energy spectrum at high wavenumbers has a universal form. Kolmogorov introduced the concept of local isotropy is to dissociate this section from the strongly anisotropic characteristics of the main flow from which is it generated. But Kolmogorov stated (1941b) that this hypothesis of local isotropy would apply to "sufficiently small domains G of the four dimensional space ($x_1, x_2, x_3, t$) *not lying near the boundary of the flow or its other singularities*". In fact spectral measurements made at all positions including near the wall all show a wave number space where there does exist a range of dissipative scales. For example the shear stress of Lawn (1991) at the wall, at the pipe axis and other radial positions are all very similar. Every known measurement near the point (11.8, 11.8) shows that the flow field is strongly anisotropic. This discrepancy can be settled with an alternative explanation. The solution of order $\varepsilon^0$ where most of the viscous dissipation occurs is independent of the solution of order $\varepsilon$ linked to the ejection process. This decoupling of the wall layer and outer flow phenomena is well demonstrated in a DNS experiment of Jimenez and Pinelli entitled "The autonomous cycle of near-wall turbulence" (1999). This dissociation of the near wall viscous diffusion and kinetic energy dominated large scale motion does not require an assumption of local isotropy.

The correlation function defines a scale often viewed as an eddy size. Its transform is seen as a distribution of kinetic energies among eddies of different sizes. The stress spectrum derived from time averaged measurements of the time averaged velocity profile in section 4.7 indicates that the wavenumber does not automatically translate to an eddy scale but can be interpreted as the distance travelled by a velocity front. The Stokes solution shows how a uniform velocity front degenerates through the penetration of viscous retardation from the wall and this can be well described by a correlation function since we are analyzing velocity behaviour within the same coherent body of fluid. The low speed streaks modelled by the solution of order $\varepsilon^0$ are well documented but the cascade of eddies implied in experimental energy spectra derived from velocity and shear measurements in the wall layer are yet to be detected in visual observations of coherent structures.

# 7  Conclusion

The classical statistics of turbulence have been shown to apply equally well to recurring unsteady state viscous flow. Care must be exercised in using them to make deductions about turbulence structures and mechanisms. It is argued that conditionally averaged statistics, particularly involving the velocities of the streaming jets in the burst phase, are more distinctive of turbulence.